\documentclass[aps,prb,twocolumn,groupedaddress]{revtex4-1}

\usepackage[utf8]{inputenc}
\usepackage{graphicx}
\usepackage{bm}
\usepackage{amsmath}
\usepackage{amssymb}
\usepackage{siunitx}
\usepackage{mathtools}
\usepackage{hyperref}
\usepackage{physics}
\hypersetup{colorlinks=true,urlcolor=black,linkcolor=black,citecolor=black}

\newcommand\plot[3]{\begin{figure} \includegraphics[#2]{plot_#1}\caption{\label{fig:#1} #3}\end{figure}}
\newcommand\fig[3]{\begin{figure} \includegraphics[#2]{fig_#1}\caption{\label{fig:#1} #3}\end{figure}}
\newcommand\mat[1]{\bm{#1}}
\newcommand\vek[1]{\bm{#1}}
\newcommand\matrdd[4]{\begin{pmatrix} #1 & #2 \\ #3 & #4 \end{pmatrix}}

\newcommand\defeq{\vcentcolon =}

\newcommand\ladd{^{\phantom{\dag}}}
\newcommand\ecom{\,,}
\newcommand\edot{\,.}
\newcommand\ead[1]{\mathrm{e}^{#1}}
\newcommand\D{\mathop{}\!\mathrm{d}}
\newcommand{\gvek}[1]{\bm{#1}}
\newcommand{\rvek}[1]{\mathcal{#1}}
\newcommand\pdim{\mathcal{P}}
\newcommand\xdim{\mathcal{X}}
\newcommand\qdim{\mathcal{Q}}
\newcommand\edim{\mathcal{E}}
\newcommand\mdim{\mathcal{M}}

\newcommand\ddim{\mathcal{D}}

\begin{document}

\title{Magnetoresistance of a Three-dimensional Dirac Gas}



\author{Viktor Könye}
\author{Masao Ogata}
\affiliation{Department of Physics, The University of Tokyo, Bunkyo-ku, Tokyo 113-0033, Japan}

\date{\today}

\begin{abstract}
We study the transversal magnetoconductivity and magnetoresistance of a massive Dirac fermion gas. This can be used as a simple model for gapped Dirac materials. In the zero mass limit the case of gapless Dirac semimetals is also studied. In the case of Weyl semimetals, to reproduce the non-saturating linear magnetoresistance seen in experiments, the use of screened charged impurities is inevitable. In this paper these are included using the first Born approximation for the self-energy. The screening wavenumber is calculated using the random phase approximation with the polarization function taking into account the electron-electron interaction. The Hall conductivity is calculated analytically in the case of no impurities and is shown to be perfectly inversely proportional to the magnetic field. Thus, the magnetic field dependence of the magnetoresistance is mainly determined by $\sigma_{xx}$. We show that in the extreme quantum limit at very high magnetic fields the gapped Dirac materials are expected to have $\sigma_{xx}\propto B^{-3}$ leading to $\varrho_{xx}\propto B^{-1}$, in contrast with the gapless case where $\sigma_{xx}\propto B^{-1}$ and $\varrho_{xx}\propto B$. At lower fields we find that the effect of the mass term is negligible and in the region of the Shubnikov-de Haas oscillations the two systems behave almost identically. We suggest a phenomenological scattering rate that is able to reproduce the linear behavior at the oscillating region. We show that in the case of the scattering rate calculated using the Born approximation the strength of the relative permittivity and the density of impurities affects the magnetic field dependence of the conductivity significantly.
\end{abstract}


\maketitle

\section{Introduction}

After the isolation of graphene in 2004\cite{Novoselov2004} the study of massless Dirac fermions in condensed matter systems became prominent over the last few decades. Following the theoretical and experimental discovery of several other two-dimensional massless fermions\cite{Wang2015}, they were also theoretically proposed in three dimensions\cite{Young2012,Wang2013a} and later found experimentally. Three-dimensional Dirac materials studied extensively are for example $\mathrm{Cd}_3\mathrm{As}_2$\cite{Liu2014,Borisenko2014}, $\mathrm{Na}_3\mathrm{Bi}$\cite{Liu2014a} and $\mathrm{TaAs}$\cite{Xu2015}. These materials are topological phases of matter and are classified as Dirac or Weyl semimetals\cite{Chiu2016}. The simplest model to describe a single independent Weyl node is the $H=\vek{\sigma}\vek{p}$ Weyl Hamiltonian \cite{Armitage2018}. In the case of Dirac semimetals there are two degenerated Weyl nodes and the Hamiltonian becomes the $4\times4$ Dirac Hamiltonian.

Recently, gapped Dirac semimetals have attracted a lot of attention, since they are expected to have very valuable applications in advanced electronic devices\cite{Zhu2017,Song2016}. Experimental realizations were found both in two dimensions\cite{Hunt2013,Ye2018} and three dimensions\cite{Zhu2017,Chen2017}. A simple effective model for these materials is the general $4\times 4$ Dirac Hamiltonian with a finite mass term\cite{Kariyado2011,Kariyado2012,Kariyado2017}. The dispersion relation is equivalent to that of relativistic fermions, but with effective values for the mass of the fermion and speed of light. In the limit of zero mass term the excitations become massless Dirac fermions as in Dirac semimetals.

Three-dimensional Dirac materials show a lot of exotic phenomena that are not present in usual systems nor in two-dimensional Dirac systems. One of these interesting features is the chiral anomaly and as a consequence negative magnetoresistance in Weyl semimetals. In Weyl semimetals Weyl nodes come in pairs with opposite chirality. In a magnetic field parallel to the electric field the chiral symmetry is broken leading to the chiral anomaly\cite{Nielsen1983} (also called Adler-Bell-Jackiw anomaly). In transport measurements this leads to a negative longitudinal magnetoresistance\cite{Huang2015,Niemann2017}. A very interesting and unique feature that seems to be present in all three-dimensional gapless Dirac materials is a non-saturating linear transverse magnetoresistance\cite{Liang2015,Narayanan2015,Niemann2017,He2014,Feng2015}. The present paper focuses on the theoretical description of this phenomenon. At the moment a complete consistent understanding is still not available for this effect. As a generalization we also study the effects of a finite mass term, how does it change the magnetoresistance and how robust is the linear behavior against the gap opening.

The transverse magnetoconductivity and magnetoresistance of Dirac materials are widely studied both in two-dimensional\cite{Shon1998a} and three-dimensional\cite{Canuto1970,Kaminker1981,Abrikosov1998,Klier2015a,Xiao2017,Klier2017a,Wang2018,Suetsugu2018} materials.  The more recent calculations are carried out using the Kubo formula and either the first Born or the self-consistent Born approximation\cite{Bruus2004}. 

A formalism that was able to describe the linear behavior for Weyl semimetals was proposed by Abrikosov\cite{Abrikosov1998}. He used the first Born approximation to calculate the scattering rate and the Kubo formula for the conductivity. The important assumption that led to his result was that the impurity potential is a screened Coulomb potential. He only studied the case with zero chemical potential and at very high magnetic fields where only the zeroth Landau level contributes to the conductivity. More recent studies\cite{Xiao2017,Klier2015a,Klier2017a} revisited this calculation in more detail. It was shown that the screened Coulomb impurities used by Abrikosov are crucial for reproducing the linear behavior\cite{Klier2015a}, since a completely different behavior is achieved for the simple case of short-range scatterers. In Ref. \onlinecite{Xiao2017} Xiao et al. calculated the scattering rates for different Landau levels, and they showed that there is Landau level dependence of the scattering rate. In their result they recovered the linear magnetoresistance for high magnetic fields, but at low fields they obtained a $B^{1/3}$ behavior. Also the effect of the first Landau level is very strong and gives a strong jump in the magnetoresistance. Klier et al.\cite{Klier2015a,Klier2017a} gave an analytic argument using the self-consistent Born approximation and several approximations. They determined the scaling of the conductivity in the different magnetic field regimes. For the high field they recovered the linear behavior. For the low fields they got $B^{4/3}$ behavior. This is not consistent with the result of Ref. \onlinecite{Xiao2017} and the inconsistency is not yet understood.

However, the above studies only discussed the massless case. In the case of massive fermions, some older references\cite{Canuto1970,Kaminker1981} discussed the problem in the context of astrophysics and thus the formalism and approximations are not exactly applicable for solid state systems. On the other hand, a recent study of the transverse magnetoresistance in gapped Dirac semimetals only used short-range scatterers and a very simple model for the scattering rate\cite{Wang2018}. In the case of longitudinal magnetoresistance the self-consistent Born approximation was discussed in the case of short-range impurities in Ref. \onlinecite{Proskurin2015}. However, as shown above, the choice of impurity potential is crucial. Therefore, in order to have a proper description for the massive Dirac fermions, the inclusion of the screening is inevitable.

In this paper, we calculate the magnetoconductivity and magnetoresistivity using the first Born approximation for the massive Dirac Hamiltonian and assuming screened charged impurities. The screening is calculated taking into account the electron-electron interaction through the polarization diagram using the random phase approximation (bubble diagram). For the massless case our result is consistent to that in Ref. \onlinecite{Xiao2017}. In the massless case we study the effects of different scattering rate choices phenomenologically, and give a scenario where the linear behavior is recovered at low fields. We calculate in detail the case of massive Dirac materials and show that the behavior is very different from the massless case at high magnetic fields.

\section{Model}

  We study a three-dimensional relativistic electron gas in a constant magnetic field. The one particle dynamics is described through the Dirac equation and the one particle Hamiltonian is:
  \begin{equation}
    \label{eq:ham}
    \mat{H}\defeq\mat{\gamma}^0\left[\sum\limits_{i=1}^3v\mat{\gamma}^i\left(p_i+eA_i\right)+\Delta\right],
  \end{equation}
  where $v$ replaces $c$ and $\Delta$ replaces $mc^2$ in the usual Dirac Hamiltonian. For the Dirac matrices ($\mat{\gamma}^\mu$) the usual Dirac representation will be used. The external uniform magnetic field is assumed to point in the $z$ direction, and the Landau gauge $\vek{A}=(0,Bx,0)$ is used. From now on $v=1$ and $\hbar=1$ is used without losing generality. With these the Hamiltonian can be expressed as:
  \begin{equation}
    \mat{H} = \begin{pmatrix} 
		\Delta 		& 0	 	& \pi_z 	& \pi_x-i\pi_y  \\
		0 		& \Delta 	& \pi_x+i\pi_y	& -\pi_z	\\
		\pi_z 		& \pi_x-i\pi_y  & -\Delta	& 0		\\
                \pi_x+i\pi_y 	& -\pi_z	& 0		& -\Delta
	      \end{pmatrix}\ecom
  \end{equation}
  where $\pi_i\defeq p_i+eA_i$. We can define the following bosonic ladder operators\cite{Ashby2013} that satisfy $[a,a^\dag]=1$:
  \begin{align}
    a&\defeq\frac{\ell_B}{\sqrt{2}}(\pi_x- i\pi_y)\ecom & a^\dag&\defeq\frac{\ell_B}{\sqrt{2}}(\pi_x- i\pi_y) \ecom
  \end{align}
  where $\ell_B\defeq\sqrt{1/eB}$ is the magnetic length. This length will be used as a natural length scale in the following (At $B=\SI{1}{\tesla}$, $\ell_B\approx\SI{25.66}{\nano\metre}$). Using these the Hamiltonian can be expressed as:
  \begin{equation}
    \mat{H} = \begin{pmatrix} 
		\Delta 			& 0	 		& p_z 			& \frac{\sqrt{2}}{\ell_B} a  \\
		0	 		& \Delta 		& \frac{\sqrt{2}}{\ell_B} a^\dag	& -p_z	\\
		p_z 			& \frac{\sqrt{2}}{\ell_B} a& -\Delta		& 0		\\
                \frac{\sqrt{2}}{\ell_B} a^\dag 	& -p_z			& 0			& -\Delta
	      \end{pmatrix}\edot
  \end{equation}

This can be solved using the eigenstates $\ket{n}$ of the $a^\dag a$ operator ($a^\dag a\ket{n}=n\ket{n}$). The energy eigenvalues and thus the Landau levels are given by:
  \begin{equation}
    \label{eq:landau}
    E_{n\lambda s}(p_z)=\lambda\sqrt{2neB+\Delta^2+p_z^2}\ecom
  \end{equation}
where $n=0,1,2,\dots$ is the Landau index, $\lambda=\pm1$ represents the band index and $s=\pm1$ represents the two-fold degeneracy (for $n\neq0$ levels). The obtained Landau levels are shown on Fig. \ref{fig:landau_levels}. Compared to the Weyl Hamiltonian\cite{Abrikosov1998,Ashby2013} the main difference is the gap present in the energy spectrum. The $n=0$ Landau level is no longer completely linear as in the case of $\Delta=0$.

    \plot{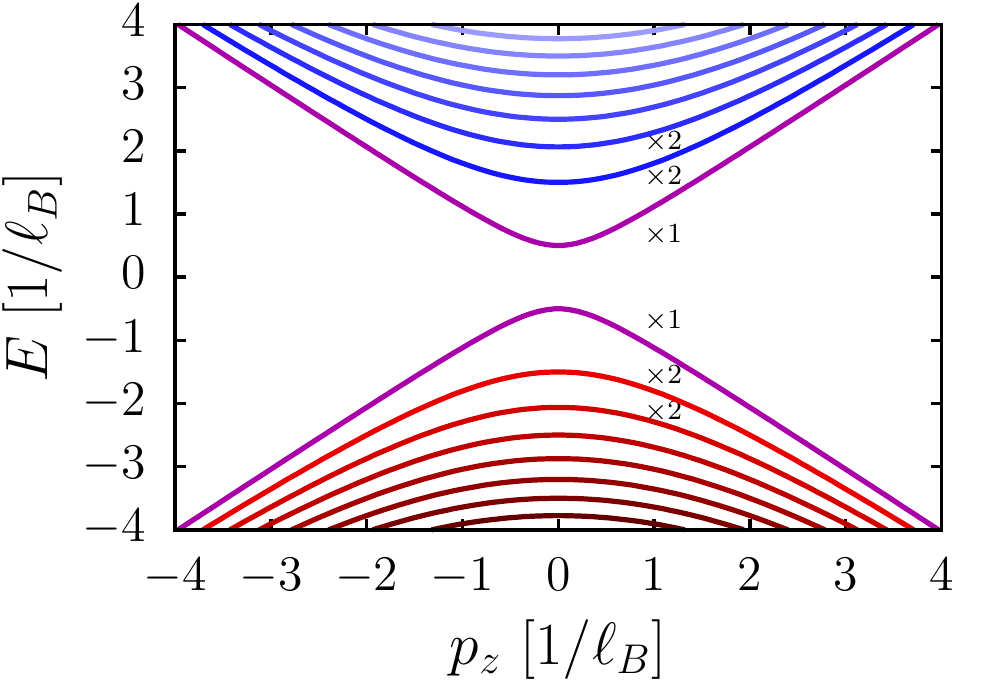}{width=.45\textwidth}{Landau levels (\ref{eq:landau}) of the Dirac Hamiltonian. The degeneracy of each level is 2 except the $n=0$ levels. $\Delta=0.5/\ell_B$ and $\ell_B=\sqrt{1/eB}$ are used.}
The eigenstates are\cite{Kaminker1981}:
  \begin{equation}
    \label{eq:eigs1}
    \ket{\vek{\Phi}_{n\lambda s}} = \begin{pmatrix*}[l] \phantom{-\lambda} u_{n,\lambda,s}&\ket{n-1} \\ \phantom{-\lambda}\llap{$-s$}u_{n,\lambda,-s}&\ket{n} \\ \phantom{-\lambda}\llap{$s\lambda$} u_{n,-\lambda,s}&\ket{n-1} \\ -\lambda u_{n,-\lambda,-s}&\ket{n}  \end{pmatrix*}
  \end{equation}
for $n\neq0$ and
  \begin{equation}
    \label{eq:eigs2}
    \ket{\vek{\Phi}_{0\lambda}} = \begin{pmatrix*}[l] \phantom{-\lambda} 0& \\ \phantom{-\lambda}\llap{$-\tilde{s}$}u_{n,\lambda,-\tilde{s}}&\ket{0} \\ \phantom{-\lambda} 0& \\ -\lambda u_{n,-\lambda,-\tilde{s}}&\ket{0}  \end{pmatrix*}
  \end{equation}
for $n=0$ where $\tilde{s}=-\mathrm{sgn}(p_z)$ and $u_{n\lambda s}$ is given by:
  \begin{equation}
    \label{eq:un}
    u_{n\lambda s}=\frac{1}{2}\sqrt{\left( 1+\frac{sp_z}{\sqrt{E_n^2-\Delta^2}}   \right)\left( 1+\lambda\frac{\Delta}{E_n}    \right)}\ecom
  \end{equation}
with $E_n\equiv E_{n11}(p_z)$. The quantum numbers describing these states are $a\equiv(n,\lambda,s,p_z,p_y)$. The dispersion relation only depends on $n$, $\lambda$ and $p_z$. Each Landau level is $L^2/2\pi\ell_B^2$-fold degenerate in $p_y$ ($L$ is the length of the system) and twofold degenerate in $s$ (for $n\neq0$). The $n=0$ Landau level must be treated with caution since there is no twofold degeneracy in $s$. The wave function of the state $\ket{n}$ can be expressed with the orthonormal Hermite-functions:
  \begin{equation}
  h_n(x;\ell_B) \defeq \frac{(\ell_B^2\pi)^{-1/4}}{\sqrt{2^nn!}}\exp(-\frac{x^2}{2\ell_B^2})H_n\left(\frac{x}{\ell_B}\right)\ecom
  \end{equation}
where $H_n(x)$ are the Hermite-polynomials. With these the eigenfunctions are:
  \begin{equation}
    \braket{x}{n} = \frac{i^n}{L}h_n(x+\ell_B^2p_y;\ell_B)\ead{ip_y y}\ead{ip_z z}\edot
  \end{equation}

Using the eigenstates $\ket{\vek{\Phi}_a}$ of the Hamiltonian, the Matsubara Green's function can be expressed as\cite{Bruus2004}:
\begin{equation}
\mat{G}^{(0)}(i\omega_m) = \sum\limits_{a}\frac{\ket{\gvek{\Phi}_{a}}\bra{\gvek{\Phi}_{a}}}{i\omega_m+\mu-E_{a}}\edot
\end{equation}
For practical reasons we will use several representations in the following. Using the wave functions defined as $\vek{\phi}_a(\vek{x})\defeq\braket{\vek{x}}{\vek{\Phi}_a}$, the non-interacting Green's function in the coordinate representation becomes:
\begin{equation}
\mat{G}^{(0)}(\vek{x},\vek{x}',i\omega_m) = \sum\limits_{a}\frac{\gvek{\phi}_{a}\ladd(\vek{x})\gvek{\phi}_{a}^\dag(\vek{x}')}{i\omega_m+\mu-E_{a}}\edot
\end{equation}
Later, the impurity averaging will be carried out in the momentum representation given by:
\begin{equation}
\mat{G}^{(0)}_{\vek{k}\vek{k}'}(i\omega_m) = \int\D^3 x \D^3 x' \ead{-i\vek{k}\vek{x}}\mat{G}^{(0)}(\vek{x},\vek{x}',i\omega_m)\ead{i\vek{k}'\vek{x}'}\ecom
\end{equation}
\begin{equation}
\mat{G}^{(0)}(x,x',i\omega_m) = \frac{1}{V^2}\sum\limits_{\vek{k},\vek{k}'}\ead{i\vek{kx}}\mat{G}^{(0)}_{\vek{kk}'}(i\omega_m)\ead{-i\vek{k}'\vek{x}'}\ecom
\end{equation}
with $V=L^3$. This $\mat{G}^{(0)}_{\vek{k}\vek{k}'}(i\omega_m)$ can be expressed using the Fourier transformed wave functions $\vek{\phi}_a(\vek{k})$ as:
\begin{equation}
\mat{G}^{(0)}_{\vek{k}\vek{k}'}(i\omega_m) = \sum\limits_{a}\frac{\gvek{\phi}_{a}\ladd(\vek{k})\gvek{\phi}_{a}^\dag(\vek{k}')}{i\omega_m+\mu-E_{a}}~.
\end{equation}

It is important to note here that in usual systems the Green's function is diagonal in the momentum space, but in the current case the position dependence of the vector potential in the Hamiltonian breaks the translational invariance thus the diagonality in the momentum space is not true. Although for $k_y$ and $k_z$ the diagonality still holds, we keep both $\vek{k}$ and $\vek{k}'$ for the sake of simplicity and generality.

The final representation is the Landau level representation where the Green's function is expressed using the eigenstates in Eqs. (\ref{eq:eigs1}) and (\ref{eq:eigs2}):
\begin{align}
G_{ba} &= \sum\limits_{\vek{k},\vek{k}'}\gvek{\phi}_b^\dag(\vek{k}) \mat{G}_{\vek{k}\vek{k}'}\gvek{\phi}_a(\vek{k}')\edot
\end{align}
Since these are the eigenstates of the Hamiltonian the non-interacting Green's function is diagonal:
\begin{equation}
G^{(0)}_{ba}(i\omega_m) = \frac{\delta_{ab}}{i\omega_m+\mu-E_a}\edot
\end{equation}

\section{Formalism}

  \subsection{Chemical potential}
The chemical potential is obtained by fixing the number density of charge carriers. Similarly to Ref. \onlinecite{Xiao2017} the density of charge carriers can be expressed as the difference of the density of electrons and holes:
\begin{equation}
\label{eq:carrdens}
n_e=\frac{1}{2\pi\ell_B^2}\int\limits_{-\infty}^\infty \frac{\D p_z}{2\pi} \sum\limits_{n=0}^\infty(2-\delta_{n0}) \left[f(E_n-\mu)-f(E_n+\mu) \right]\ecom
\end{equation}
where the factor $(2-\delta_{n0})$ is taking care of the different degeneracy of $s$ for the zeroth Landau level.
Using the density of states $D(\varepsilon)$, Eq. (\ref{eq:carrdens}) can be expressed as:
\begin{align}
\label{eq:chemdos}
n_e&=\int\limits_{0}^\infty\D \varepsilon D(\varepsilon)\left[f(\varepsilon-\mu)-f(\varepsilon+\mu) \right]\ecom\\
\label{eq:dos}
D(\varepsilon)&=\frac{1}{2\pi\ell_B^2}\int\limits_{-\infty}^\infty \frac{\D p_z}{2\pi} \sum\limits_{n=0}^\infty(2-\delta_{n0})\left[\delta(\varepsilon-E_n)+\delta(\varepsilon+E_n)\right]\edot
\end{align}
In Eq. (\ref{eq:chemdos}) we have used $D(\varepsilon)=D(-\varepsilon)$. In the following the chemical potential is calculated implicitly solving one of the above equations.
  \subsection{Self-energy}
  \label{sec:selfen}
The Green's function with impurities will be approximated using the first-order Born approximation\cite{Bruus2004}. The self energy is obtained using the Feynman diagram shown in Fig. \ref{fig:feyn_born} similarly to previous studies\cite{Abrikosov1998,Xiao2017}:
  \begin{equation}
    \label{eq:born}
    \mat{\Sigma}^B_{\vek{k}\vek{k}'}(i\omega_m) = n_i\frac{1}{V}\sum\limits_{\vek{q}}u_{\vek{q}}^2\mat{G}^{(0)}_{\vek{k-q},\vek{k'-q}}(i\omega_m)\ecom
  \end{equation}
where $n_i$ is the number density of the impurities and $u_{\vek{q}}$ is the Fourier transform of the effective impurity potential.
\fig{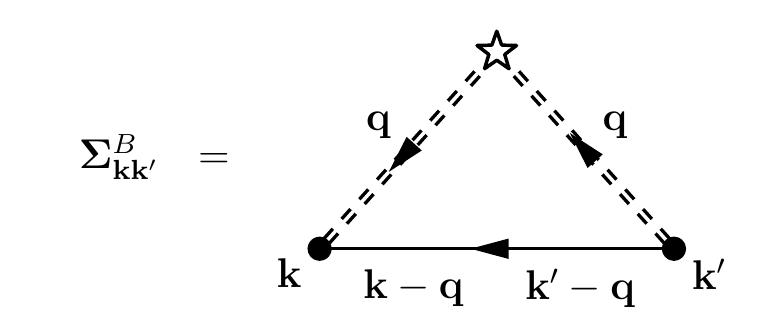}{width=.45\textwidth}{Feynman diagram for the first-order Born approximation of the self-energy (see Eq. (\ref{eq:born})).}
Since the translational invariance is broken the self-energy is not diagonal in the momentum space. The self-energy in the Landau level representation is calculated as:
  \begin{equation}
    \label{eq:selfen}
    \Sigma_{ba}(i\omega_m) = \sum\limits_{\vek{k},\vek{k}'}\gvek{\phi}_b^\dag(\vek{k}) \mat{\Sigma}_{\vek{k}\vek{k}'}(i\omega_m)\gvek{\phi}_a(\vek{k}')\edot
  \end{equation}

Using the Landau level representation of the self-energy the Dyson equation is (at $i\omega_m$ frequency):
  \begin{equation}
    G_{ab} = \delta_{ab}G^{(0)}_a + G^{(0)}_a\sum\limits_{c}\Sigma_{ac}G_{cb}\edot
  \end{equation}
At this point we will assume that the self-energy is diagonal in the Landau level representation. This is not proven analytically, but checking several non-diagonal elements numerically we find that the difference between diagonal and non diagonal elements is several orders of magnitude (in the magnetic field ranges used in following sections), thus the diagonality is a valid assumption. Also we will assume that the real part of the self energy is renormalized into the chemical potential and only use the scattering rate $\Gamma_a=-\Im{\Sigma_a}$. With these, the Green's function becomes diagonal and using the Dyson equation it is simply given as:
  \begin{equation}
    \label{eq:green}
   G_a(i\omega_m) = \frac{1}{i\omega_m+\mu-E_a+i\Gamma_a(i\omega_m)}\edot
  \end{equation}

  \subsection{Impurity potential}
The effective impurity potential $u_{\vek{q}}$ will be calculated using the so called Random Phase Approximation\cite{Bruus2004,Mahan2000} (RPA). Since the Green's function is not diagonal in the momentum space, the treatment of the RPA must be performed with caution. Diagrammatically the screened impurity potential is expressed as in Fig. \ref{fig:feyn_screen}. 
\fig{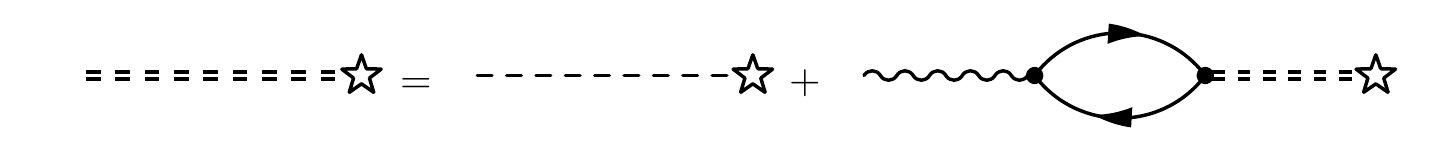}{width=.45\textwidth}{Feynman diagram for the RPA of the impurity potential. The double dashed lines are the effective impurity potentials, the single dashed lines are the bare impurity potentials and the wavy line represents the electron-electron interaction.}
Starting from a charged impurity in a dielectric medium $v_{\vek{q}}=u_i/q^2$ (single dashed line) and the electron-electron interaction $w_{\vek{q}}=u_e/q^2$ (wavy line) the screened impurity potential is expressed implicitly as:
  \begin{equation}
    u_{\vek{q}\vek{q}'}(i\omega_\lambda) = v_{\vek{q}}\delta_{\vek{q}\vek{q}'}+\frac{1}{V}\sum\limits_{\vek{q}''}w_{\vek{q}}\Pi^0_{\vek{q}\vek{q}''}(i\omega_\lambda)u_{\vek{q}''\vek{q}'}(i\omega_\lambda)\ecom
  \end{equation}
where $\Pi^0$ is the bubble diagram for electrons. For simplicity we study the static and long-wave limit $i\omega_\lambda,\vek{q},\vek{q}'\to 0$. In this limit we assume that the bubble diagram is diagonal and constant $\Pi^0_{\vek{qq}'}(i\omega_m)\approx\delta_{\vek{qq}'}\Pi^0_{\vek{00}}(0)$. With this the effective screening will also be diagonal and simply expressed as:
  \begin{equation}
    u_{\vek{q}}=\frac{u_i}{q^2-u_e\Pi^0_{\vek{00}}(0)}\equiv\frac{u_i}{q^2+\kappa^2}\edot
  \end{equation}
\fig{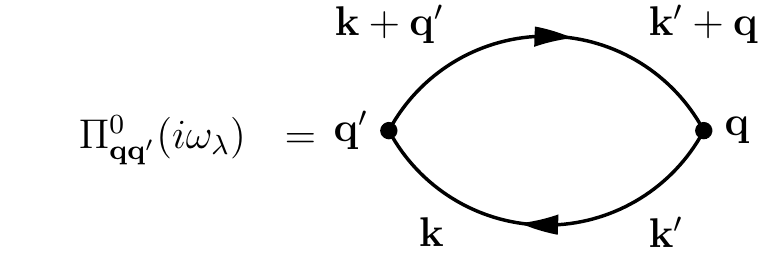}{width=.45\textwidth}{The bubble diagram used to calculate the screening wavenumber in Eq. (\ref{eq:kappa}).}
Here $\kappa$ is the screening wavenumber and it is obtained using the static and long-wave limit of the bubble diagram (see Fig. \ref{fig:feyn_bubble}) as:
  \begin{equation}
    \label{eq:kappa}
    \kappa^2=-\frac{u_e}{\beta V^2}\sum\limits_{m,\vek{k},\vek{k}'}\Tr\left[\mat{G}_{\vek{kk}'}(i\omega_m)\mat{G}_{\vek{k}'\vek{k}}(i\omega_m)\right]\edot
  \end{equation}
  \subsection{Linear response theory}
  \label{sec:lrt}
The one particle current operator of the system is:
  \begin{equation}
    \label{eq:curr}
    \mat{J}_i = \mat{\gamma}^0\mat{\gamma}^i=\matrdd{0}{\mat{\sigma}^i}{\mat{\sigma}^i}{0}\edot
  \end{equation}
In the Landau level representation, this is:
  \begin{equation}
    J^{(i)}_{ab} = \int\D^3 x \gvek{\phi}_a^\dag(\vek{x}) \mat{J}_i \gvek{\phi}_b(\vek{x})\edot
  \end{equation}
The optical conductivity is calculated using the Matsubara current-current correlation function\cite{Bruus2004} as:
    \begin{equation}
\sigma_{ij}(\omega) = \frac{ie^2}{\omega} \lim\limits_{\delta\to0^+}\Pi_{ij}(i\omega_\lambda=\omega+i\delta)\ecom
  \end{equation}
where the correlation function is calculated as:
  \begin{align}
    \Pi_{ij}(i\omega_\lambda) &= \frac{1}{V}  \sum\limits_{a,b}\frac{1}{\beta}\sum\limits_n  J^{(i)}_{ab} G_b(i\omega_n+i\omega_\lambda) J^{(j)}_{ba} G_a(i\omega_n)\ecom
  \end{align}
with the Green's functions taken from Eq. (\ref{eq:green}). The vertex correction is neglected in this formula. Based on the results obtained for the Weyl Hamiltonian in Ref. \onlinecite{Klier2015a} we assume that the effect of the vertex correction is a magnetic field independent renormalization of the scattering rate. The Matsubara sum can be transformed into an integral\cite{Abrikosov1965} (keeping in mind that due to $\Gamma\propto\mathrm{sgn}(\omega_m)$ there is a branch cut):
  \begin{align}
    \Pi_{ij}(\omega) &= \frac{1}{V}  \sum\limits_{a,b} J^{(i)}_{ab} J^{(j)}_{ba} C_{ba}(\omega)\ecom
  \end{align}
where
  \begin{align}
    \nonumber
    C_{ba}(\omega)= -2&\int\limits_{-\infty}^\infty \frac{\D\varepsilon}{2\pi}\bigg[f(\varepsilon)G_b^R(\varepsilon+\omega)\Im{G^R_a(\varepsilon)}+\\
      &+f(\varepsilon+\omega)\Im{G^R_b(\varepsilon+\omega)}G^A_a(\varepsilon)\bigg]\ecom
  \end{align}
where $f(\varepsilon)$ is the Fermi-Dirac distribution and
  \begin{align}
    G^{R/A}_a(\varepsilon)&\defeq\frac{1}{\varepsilon-E_a+\mu\pm i\Gamma_a(\varepsilon)}\edot
  \end{align}

The DC conductivity is calculated by taking $\omega\to 0$:
  \begin{equation}
    \label{eq:sigma}
    \sigma_{ij} = -\lim\limits_{\omega\to0}\frac{e^2}{\omega}\Im{\Pi_{ij}(\omega)}\edot
  \end{equation}

\section{Results}

  \subsection{Chemical potential}
As a realistic charge carrier density, $n_e=\SI{1e18}{\per\cubic\centi\metre}$ will be used\cite{Liang2015}. The density of states is calculated using Eq. (\ref{eq:dos}) as:
  \begin{equation}
    \label{eq:dosres}
    D(\varepsilon)=\frac{1}{2\pi^2\ell_B^2}\hspace{-5pt}\sum\limits_{n=0}^{\left\lfloor\frac{(\varepsilon^2-\Delta^2)\ell_B^2}{2}\right\rfloor}\hspace{-5pt}(2-\delta_{n0})\frac{|\varepsilon|}{\sqrt{\varepsilon^2-\Delta^2-\frac{2n}{\ell_B^2}}}\edot
  \end{equation}
In the case of $\Delta=0$ we recover the result obtained in Ref. \onlinecite{Ashby2014}. The density of states is shown in Fig. \ref{fig:dos}.
\plot{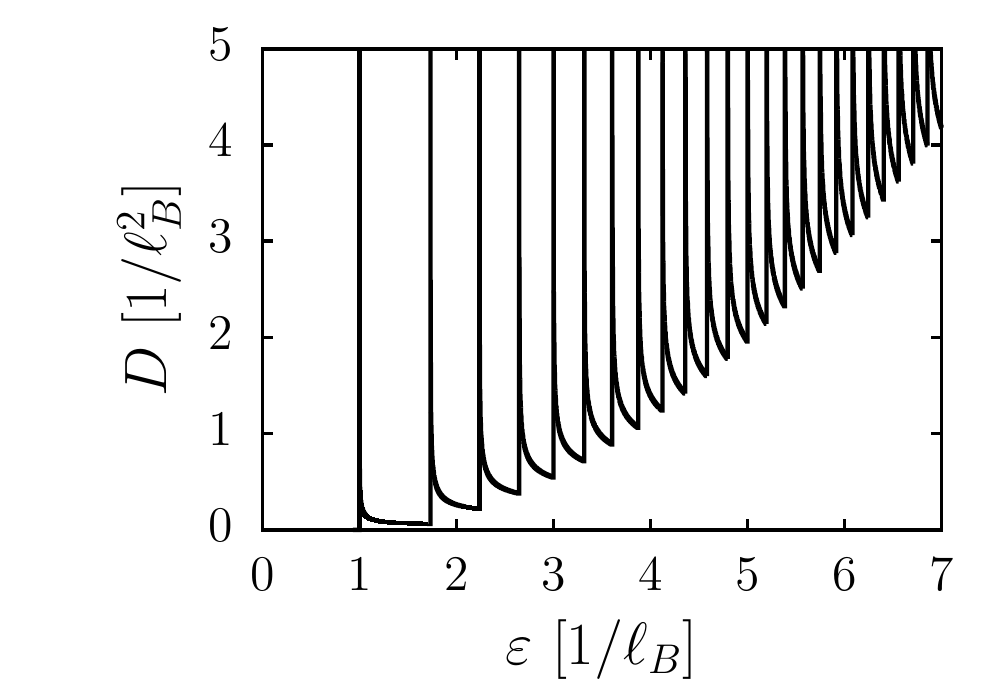}{width=.45\textwidth}{Density of states calculated from Eq. (\ref{eq:dosres}). The mass term is chosen as $\ell_B\Delta=1$.}
Substituting this in Eq. (\ref{eq:chemdos}) we obtain the expression for the charge carrier density. At zero temperature after integration:
  \begin{equation}
    \label{eq:chempotsimp}
    n_e=\frac{1}{2\pi^2\ell_B^2}\sum\limits_{n=0}^{\left\lfloor\frac{(\mu^2-\Delta^2)\ell_B^2}{2}\right\rfloor}(2-\delta_{n0})\sqrt{\mu^2-\Delta^2-\frac{2n}{\ell_B^2}}\edot
  \end{equation}
  This is consistent with the $\Delta=0$ result in Ref. \onlinecite{Xiao2017}. In the high magnetic field limit ($\ell_B\to 0$) only the zeroth Landau level contributes and the equation simply yields
  \begin{equation}
    \label{eq:chemhigh}
    \mu = \sqrt{4\pi^4n_e^2\ell_B^4+\Delta^2}\edot
  \end{equation}
  As we can see at high magnetic fields $\mu\to\Delta$, and for $\Delta=0$ $\mu\propto 1/B$ as in Ref. \onlinecite{Abrikosov1998}. In the low magnetic field limit ($\ell_B\to\infty$) the summation can be substituted with an integral and we obtain:
  \begin{equation}
    \label{eq:chemB0}
    n_e=\frac{(\mu^2-\Delta^2)^\frac{3}{2}}{3\pi^2}\ecom
  \end{equation}
which reproduces the zero magnetic field result.

\begin{figure}
      \includegraphics[width=.45\textwidth]{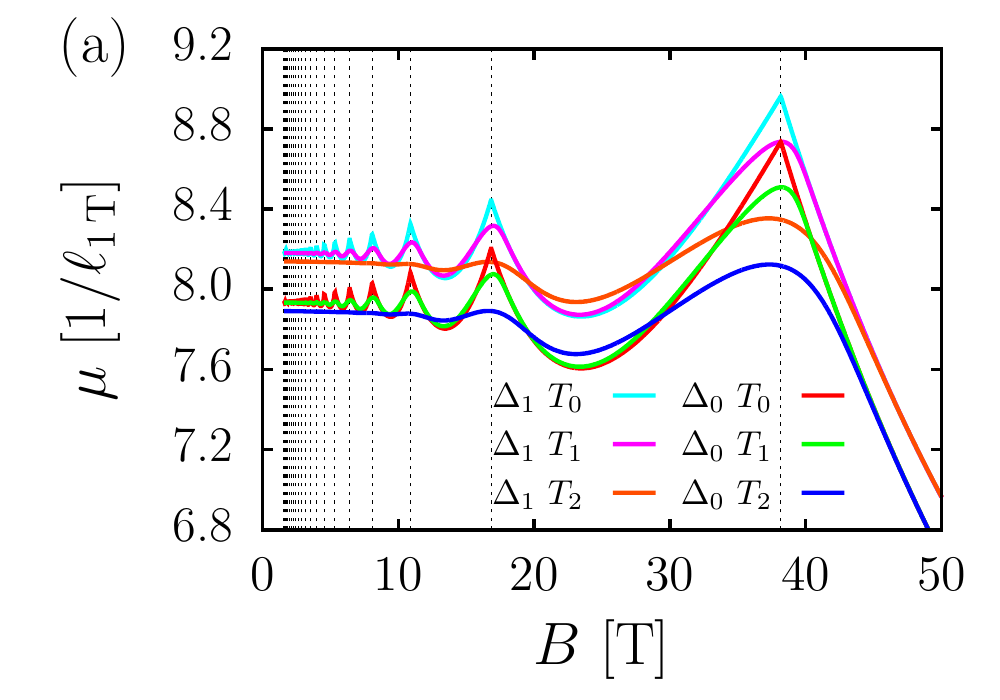}
      \includegraphics[width=.45\textwidth]{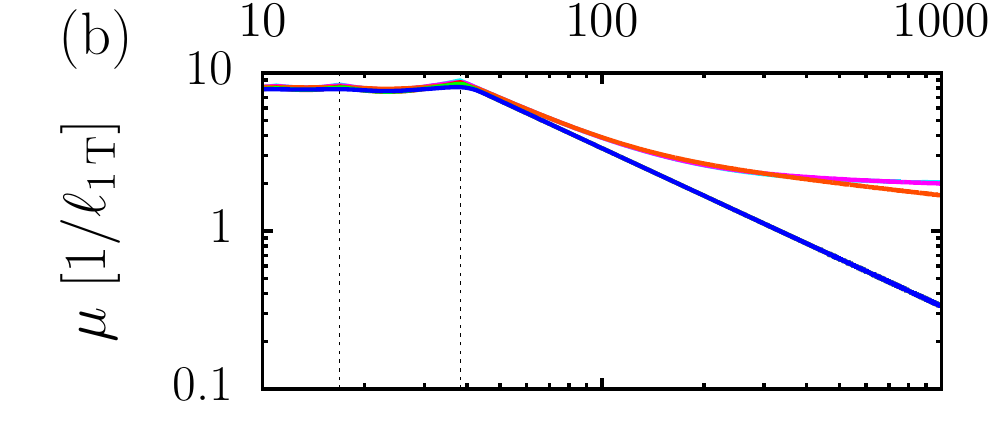}
      \caption{\label{fig:chem} Magnetic field dependence of the chemical potential for two mass terms $\Delta_0=0$ and $\Delta_1=2/\ell_{\SI{1}{\tesla}}\approx\SI{50}{\milli\electronvolt}$ and for three temperatures $T_0=\SI{0}{\kelvin}$, $T_1=\SI{30}{\kelvin}$ and $T_2=\SI{100}{\kelvin}$. The carrier density is fixed at $n_e=10^{18}\si{\per\cubic\centi\metre}$. (a) is for the small magnetic fields with oscillating behavior and (b) is for the large magnetic field limit. The vertical lines show the magnetic fields where a new Landau level crosses the chemical potential as calculated from Eq. (\ref{eq:Bpeak}).}
    \end{figure}

Solving Eq. (\ref{eq:chempotsimp}) for zero temperature or Eq. (\ref{eq:chemdos}) for finite temperature numerically we obtain the chemical potential as shown in Fig. \ref{fig:chem}.
For the high and low magnetic field limits we see the behavior explained using Eq. (\ref{eq:chempotsimp}). In the case of finite temperature we see some minor deviation but the main behavior remains the same. Between the two limits we see oscillations caused by the singularities present in the density of states. At zero temperature the oscillations are more prominent with sharp changes in the chemical potential. Finite temperature smoothens the curves and at high temperatures the oscillation almost completely disappears.

The oscillation occurs when a new Landau level crosses the chemical potential. At zero temperature the magnetic fields where the oscillation occurs can be calculated from the condition of $(\mu^2-\Delta^2)\ell_B^2/2\in \mathbb{N}$. Solving Eq. (\ref{eq:chempotsimp}) with this condition yields:
  \begin{align}
    \label{eq:Bpeak}
    B_m=\left(\frac{\sqrt{2}\pi^2n_e}{A(m)}\right)^{\frac{2}{3}},~A(m)\defeq\sum\limits_{n=0}^m(2-\delta_{n0})\sqrt{m-n}\ecom
  \end{align}
where $m\in\mathbb{N}$ denotes the $m^{\text{th}}$ Landau level that crosses the chemical potential. We can see from the formula and the numerical results as well that the peaks occur at the same magnetic field independently of $\Delta$.

  \subsection{Scattering rate}
    \label{sec:gamma}
    The screening wavenumber using Eq. (\ref{eq:kappa}) becomes:
    \begin{equation}
      \label{eq:wave}
      \kappa^2 =\frac{-u_e}{2\pi\ell_B^2}\int\limits_{-\infty}^\infty \frac{\D p_z}{2\pi} \sum\limits_{\substack{n=0\\\lambda=\pm1}}^\infty(2-\delta_{n0})\frac{\partial f(\lambda E_n-\mu)}{\partial \lambda E_n}\edot
    \end{equation}
    At zero temperature this formula becomes the same formula as the expression for the density of states (\ref{eq:dos}) evaluated at the chemical potential $\kappa^2=D(\mu)$. This can be calculated as explained in the previous section:

    \begin{equation}
      \label{eq:waveo}
      \kappa^2=\frac{u_e}{2\pi^2\ell_B^2}\sum\limits_{n=0}^{\left\lfloor\frac{(\mu^2-\Delta^2)\ell_B^2}{2}\right\rfloor}(2-\delta_{n0})\frac{|\mu|}{\sqrt{\mu^2-\Delta^2-\frac{2n}{\ell_B^2}}}\edot
    \end{equation}

    For high magnetic fields ($\ell_B\to 0$) only the zeroth Landau level contributes to the screening. Using the high magnetic field dependence of $\mu$ (\ref{eq:chemhigh}), we obtain:
    \begin{equation}
      \kappa^2\sim\frac{u_e}{2\pi^2\ell_B^2}\frac{\sqrt{4\pi^4n_e^2\ell_B^4+\Delta^2}}{2\pi^2n_e\ell_B^2}\edot
    \end{equation}
    For $B\to\infty$ $\kappa^2\propto B$ for $\Delta=0$ as used in Ref. \onlinecite{Abrikosov1998} and $\kappa^2\propto B^2$ for $\Delta\neq0$. In the zero field limit similarly to Eq. (\ref{eq:chemB0}) the sum can be substituted with an integral and
    \begin{equation}
      \kappa^2\sim\frac{u_e}{\pi^2}\sqrt{3\pi^2n_e+\left(3\pi^2n_e\right)^{\frac{1}{3}}\Delta^2}
    \end{equation}
    goes to a constant. The screening wavenumber as a function of magnetic field is shown in Fig. \ref{fig:screen}.
    \plot{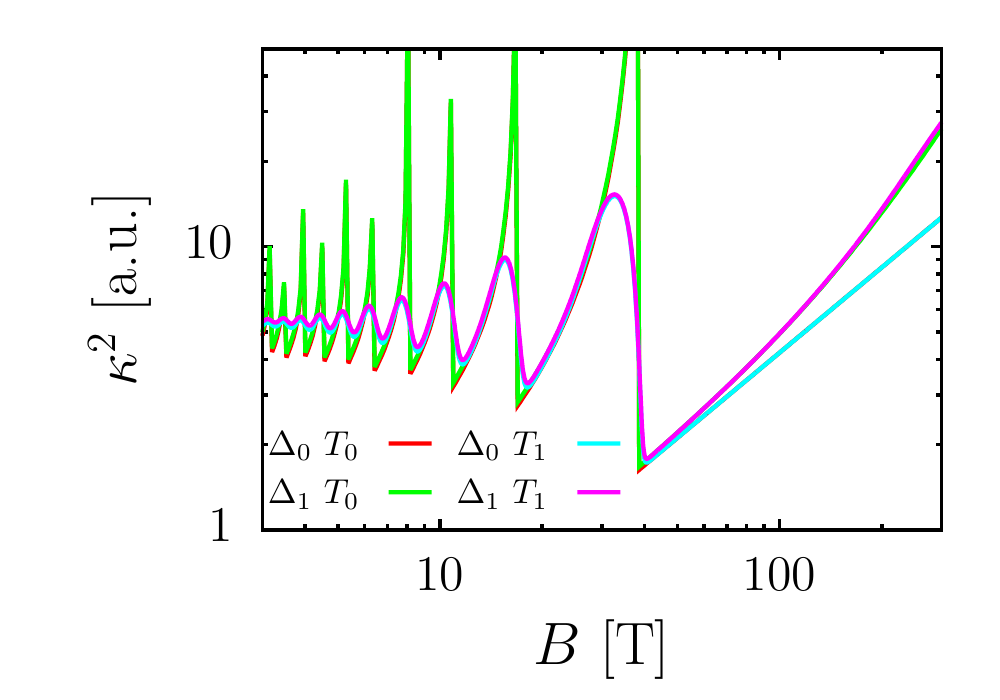}{width=.45\textwidth}{Magnetic field dependence of the screening wavenumber for several cases as in Fig. \ref{fig:chem}.}

    From now on we introduce the dimensionless quantities using $\ell_B =\sqrt{1/eB}$:
    \begin{align}
       \rvek{P} &\defeq\ell_B \vek{p}\ecom & \mathcal{E}&\defeq\ell_B\varepsilon\ecom & \mathcal{M}&\defeq\ell_B\mu\ecom & \mathcal{D}&\defeq\ell_B\Delta\edot
    \end{align}
    We have to keep in mind that $\mu$ and $\Delta$ are not integration variables but parameters thus $\mathcal{M}$ and $\mathcal{D}$ are functions of $B$. Using Eq. (\ref{eq:selfen}) for the scattering rate we obtain:
    \begin{widetext}
      \begin{subequations}
      \begin{align}
	\label{eq:gam}
        \Gamma_{n\lambda s}(\edim,\pdim_z,B) &=\frac{n_i\pi}{\ell_B^2}\sum\limits_{\ell=0}^{\left\lfloor\frac{(\edim+\mdim)^2-\mathcal{D}^2}{2} \right\rfloor}\sum\limits_{\substack{\alpha = \pm 1 \\ t=\pm1}} \int\frac{\D \qdim_x\D \qdim_y}{(2\pi)^3}u_{\vek{\qdim}_{\ell\alpha}}^2(B) \left|\frac{\edim+\mdim}{\sqrt{(\edim+\mdim)^2-2\ell-\mathcal{D}^2}} \right|
       \left|F_{n\lambda s,\ell\gamma_ot}(\vek{\qdim}_{\ell\alpha},\pdim_z)\right|^2\ecom
      \end{align}

      \begin{align}
	\label{eq:four}
        F_{n\lambda s,\ell\gamma t}(\vek{\qdim},\pdim_z) &\defeq \int\D \xdim \gvek{\phi}_{n\lambda s}^\dag(\xdim;0,\pdim_z) \gvek{\phi}_{\ell\gamma t}(\xdim;\qdim_y,\pdim_z-\qdim_z)\ead{i\qdim_x\xdim}\ecom
      \end{align}
      \begin{align}
        \vek{\qdim}_{\ell\pm} &\defeq ({\qdim_x},{\qdim_y},{\qdim_{\ell\pm}})\ecom
        &\qdim_{\ell\pm} &\defeq \pdim_z\pm\sqrt{(\edim+\mdim)^2-2\ell-\ddim^2}\edot
      \end{align}
      \end{subequations}
    \end{widetext}
where $\gamma_o = \mathrm{sgn}(\varepsilon+\mu)$ and for the impurity potential 
    \begin{equation}
        u_{\vek{\qdim}}(B) = \frac{u_i\ell_B^2}{\vek{\qdim}^2+\ell_B^2\kappa^2(B)}
    \end{equation}
is used, where the screening is calculated from Eq. (\ref{eq:wave}). Note here that the summation over $t$ is only for $l\neq0$.

    \plot{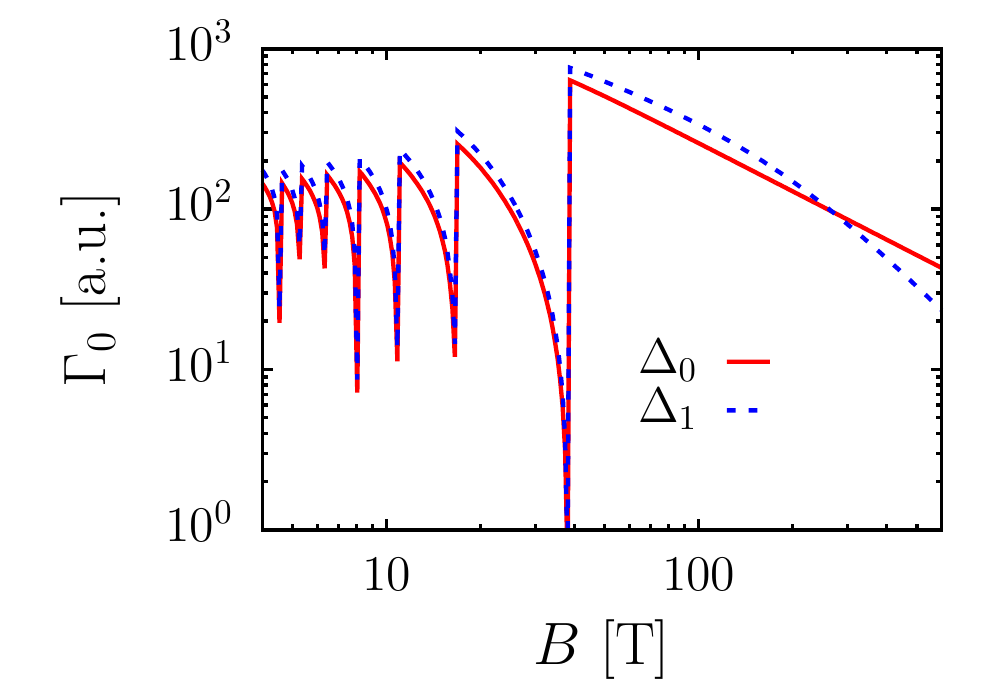}{width=.45\textwidth}{Scattering rate calculated from Eq. (\ref{eq:gam}) with indices $n=0$ and $\lambda=1$ as a function of magnetic field at zero temperature. The screening wavenumber is calculated using Eq. (\ref{eq:wave}). The density of charge carriers is $n_e=10^{18}\si{\per\cubic\centi\metre}$. $\Delta_0=0$ and $\Delta_1=2/\ell_{\SI{1}{\tesla}}$.}
    In the high magnetic field limit at $\mathcal{E}=0$ (we will see later that this is the important energy at high magnetic fields) only the zeroth Landau level plays role. Using Eq. (\ref{eq:chemhigh}) at zero temperature in the case of $\Delta=0$ gives $\Gamma\propto 1/B$ for high magnetic fields. In the case of finite mass term at high fields we get $\Gamma\propto1/B^2$. In Fig. \ref{fig:Gamma_B} we show the scattering rate as a function of the magnetic field at $\edim=0$ and $\pdim=0$. For the numerical calculations $u_e=e^2/\varepsilon_0$ is used and the energy scale is set using $v=\SI{1e6}{\metre\per\second}$ based on Refs. \onlinecite{Liang2015} and \onlinecite{Liu2014}. The Fourier transformation in Eq (\ref{eq:four}) is calculated using the fractional Fourier transform\cite{Bailey1994}. The $\qdim_x$ integral is done using the Simpson's rule on the result of the fractional Fourier transform and finally the $\qdim_y$ integral is done through Gaussian quadrature.
    \plot{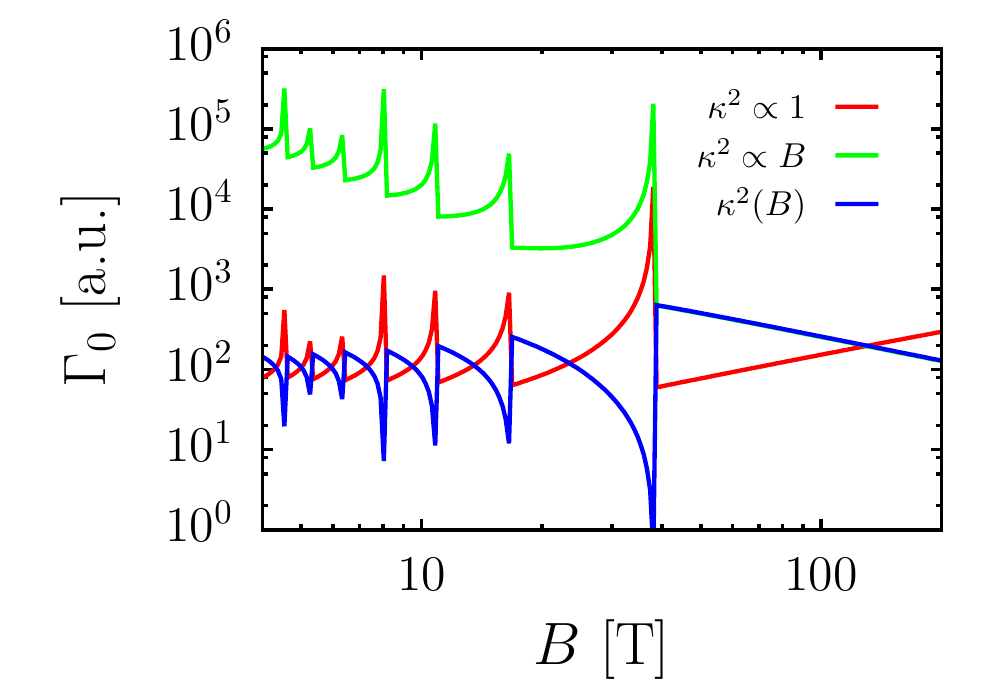}{width=.45\textwidth}{Scattering rate calculated from Eq. (\ref{eq:gam}) with indices $n=0$ and $\lambda=1$ as a function of magnetic field. The density of charge carriers is $n_e=10^{18}\si{\per\cubic\centi\metre}$. Different type of screening wavenumbers are used: a constant, a linear in magnetic field and the one calculated from Eq. (\ref{eq:waveo}). $\Delta=0$ and $T=\SI{0}{\kelvin}$.}

As shown in Fig. \ref{fig:Gamma_B}, at the high magnetic field limit we see the behavior described above. At low fields SdH oscillations can be seen. The effect of the mass term is only relevant in the extremely high magnetic field limit. As a function of the magnetic field the scattering rate first has an increasing background (with SdH oscillations) then after reaching the quantum limit it starts to decrease.

    This behavior is strongly dependent on the choice of the magnetic field dependence of the screening wavenumber. In Fig. \ref{fig:Gamma_Bs} different wavenumber choices are shown for the zero temperature, zero mass term case. If we assume that $\kappa$ is independent of $B$ (red line in Fig. \ref{fig:Gamma_Bs}) we get a monotonic increase at high magnetic fields and at lower fields there are divergent peaks with a constant background. If we assume that $\kappa^2$ is proportional to $B$ (green line), $\Gamma\propto 1/B$ at high magnetic fields and at lower fields the divergent peaks are on a decreasing background (This is the screening used in Ref. \onlinecite{Abrikosov1998}). Using our screening calculated as Eq. (\ref{eq:wave}) (blue line in Fig. \ref{fig:Gamma_Bs}) we obtain a magnetic field dependence of $\Gamma$ with a maximum at the quantum limit. We will see that this behavior of increasing then decreasing scattering rate is important to reproduce the linear magnetoresistance.

  \subsection{Hall Conductivity}
    The conductivity is calculated through the steps described in Sec. \ref{sec:lrt}. In the case of the Hall conductivity we will neglect the effect of impurities since they are expected not to have a large influence on the result\cite{Abrikosov1998,Xiao2017}.

    The matrix elements of the current operator using Eq. (\ref{eq:curr}) are:
\begin{widetext}
      \begin{subequations}
    \begin{align}
        \label{eq:jx}
        J_{ab}^{(x)} &= -\phantom{i}\delta_{p_yp_y'}\delta_{p_zp_z'}\left[\delta_{n,n'-1}(\lambda u_{n,-\lambda,-s}u_{n',\lambda',s'}+ss'\lambda'u_{n,\lambda,-s}u_{n',-\lambda',s'})+(n\leftrightarrow n',\lambda\leftrightarrow\lambda',s\leftrightarrow s')\right]\ecom\\
        J_{ab}^{(y)} &= \phantom{-}i\delta_{p_yp_y'}\delta_{p_zp_z'}\left[\delta_{n,n'-1}(\lambda u_{n,-\lambda,-s}u_{n',\lambda',s'}+ss'\lambda'u_{n,\lambda,-s}u_{n',-\lambda',s'})-(n\leftrightarrow n',\lambda\leftrightarrow\lambda',s\leftrightarrow s')\right]\edot
    \end{align}
      \end{subequations}
      In the absence of impurities, the imaginary part of the Green's function can be substituted with a Dirac delta and $\Im{G_a^R(\varepsilon)}=-i\pi\delta(\varepsilon-E_a+\mu)$. Using Eq. (\ref{eq:sigma}) and evaluating the integral and the DC limit, the formula for the Hall conductivity becomes:
      \begin{equation}
        \label{eq:hall}
        \sigma_{xy} =2\frac{\sigma_0}{\ell_B}\sum\limits_{n=0}^{\infty}\sum\limits_{\substack{\lambda,\lambda' = \pm 1 \\ s,s'=\pm1}} \int \D \pdim_z \left(\lambda u_{n,-\lambda,-s}u_{n+1,\lambda',s'}+ss'\lambda'u_{n,\lambda,-s}u_{n+1,-\lambda',s'}\right)^2 \frac{f(\lambda E_n-\mdim)-f(\lambda'E_{n+1}-\mdim)}{(\lambda E_n-\lambda' E_{n+1})^2}\ecom
      \end{equation}
      where $\sigma_0=e^2/h$ is the inverse of the von Klitzing constant. The summation over $s$ is taken only for $n\neq0$. This formula can be simplified by using the properties of the Fermi distribution ( $f(-\edim-\mdim)=1-f(\edim+\mdim)$ ), the definition of $u_{n\lambda s}$ (\ref{eq:un}) and the explicit form of $E_n$. After the summations over the $\lambda,\lambda',s,s'$ indices, we can show that the formula becomes:
      \begin{equation}
        \label{eq:hall2}
        \sigma_{xy} =\frac{\sigma_0}{\ell_B}\sum\limits_{n=0}^{\infty}\int \D \pdim_z (1+2n) \big\{\left[f(E_n-\mdim)-f(E_{n}+\mdim)\right]-\left[f(E_{n+1}-\mdim)-f(E_{n+1}+\mdim)\right]\big\}\edot
      \end{equation}
      \end{widetext}
After rearranging the summation over $n$ we can see that Eq. (\ref{eq:hall2}) is proportional to Eq. (\ref{eq:carrdens}) and the Hall conductivity can be expressed using the carrier density as:
      \begin{equation}
	\label{eq:hallsimp}
        \sigma_{xy}=\sigma_02\pi\ell_B^2n_e=\frac{en_e}{B}\edot
      \end{equation}
      \plot{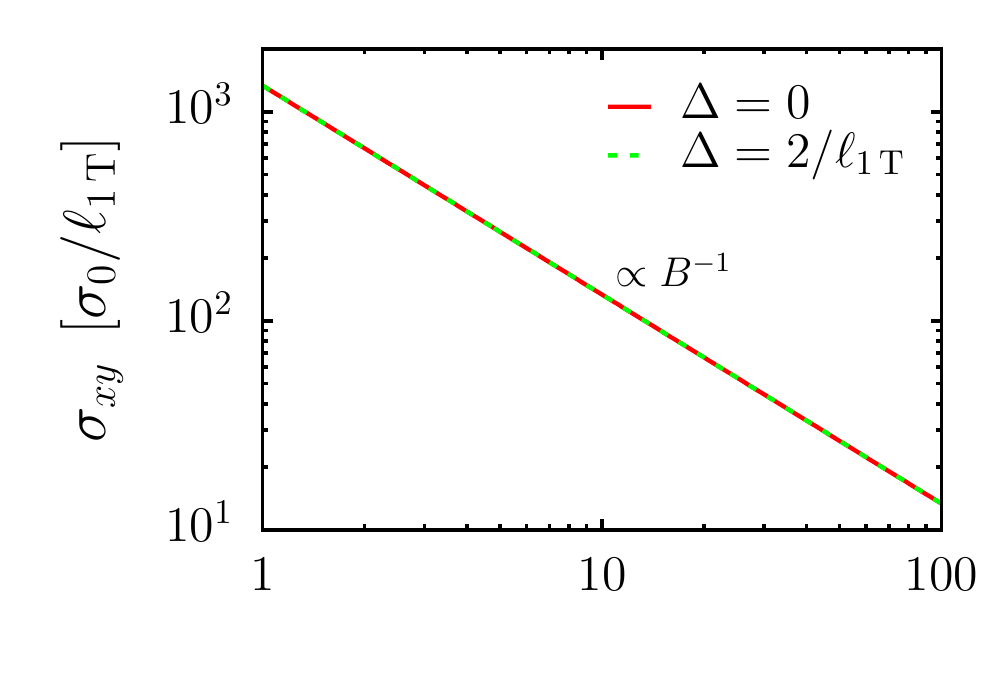}{width=.45\textwidth}{Hall conductivity $\sigma_{xy}$ calculated from Eq. (\ref{eq:hall}) as a function of magnetic field at $T=\SI{30}{\kelvin}$. The density of charge carriers is $n_e=10^{18}\si{\per\cubic\centi\metre}$. $\sigma_0/\ell_{\SI{1}{\tesla}}\approx \SI{15}{\per\ohm\per\centi\metre}$.}
Since the charge carrier density is constant the Hall conductivity is exactly inversely proportional to the magnetic field as in usual systems.
 To check the validity of Eq. (\ref{eq:hallsimp}) the Hall conductivity is calculated numerically from Eq. (\ref{eq:hall}). The numerical results at different mass terms can be seen in Fig. \ref{fig:condxy} (the results are only shown at one finite temperature, but at different temperatures we get exactly the same result).
      As we can see the Hall conductivity does not depend on the mass term nor the temperature (in the case of no impurities) and it exactly satisfies Eq. (\ref{eq:hallsimp}).

  \subsection{Diagonal conductivity}
In the case of the diagonal component including impurities is necessary in order to get finite conductivity. The impurity is included in the Green's function as explained in Sec. \ref{sec:selfen}. From Eq. (\ref{eq:jx}) we see that the matrix elements of the $x$ component of the current operator are all real. Thus, in the Eq. (\ref{eq:sigma}) when taking the imaginary part we only need the imaginary part of $C_{ba}(\omega)$. Similarly to Ref. \onlinecite{Abrikosov1998} with our notations, we obtain:
      \begin{widetext}
      \begin{subequations}
	\label{eq:sigxx}
      \begin{equation}
        \sigma_{xx} =\frac{2}{\pi}\frac{\sigma_0}{\ell_B}\sum\limits_{n=0}^{\infty}\sum\limits_{\substack{\lambda,\lambda' = \pm 1 \\ s,s'=\pm1}} \int \D \pdim_z \left(\lambda u_{n,-\lambda,-s}u_{n+1,\lambda',s'}+ss'\lambda'u_{n,\lambda,-s}u_{n+1,-\lambda',s'}\right)^2 C_{n\lambda s;n+1\lambda's'}\ecom
      \end{equation}
      \begin{equation}
        C_{ab}\defeq\frac{\beta}{4\ell_B}\int\limits_{-\infty}^\infty \frac{\D\edim}{2\pi}\frac{1}{\cosh^2\left(\frac{\beta\edim}{2\ell_B}\right)}\Im{G_a^R(\edim)}\Im{G^R_b(\edim)} \edot
      \end{equation}
      \end{subequations}
      \end{widetext}
The summation over $s$ is again only taken for $n\neq0$.

First, let us discuss analytically the behavior of $\sigma_{xx}$ in high magnetic fields. In the high magnetic field limit ($\ell_B\to0$) the formula for $C_{ab}$ becomes equivalent to the zero temperature formula since the $T\ell_B$ combination goes to zero. Thus, we have:
      \begin{equation}
        C_{ab}=\frac{1}{2\pi}\Im{G_a^R(0)}\Im{G^R_b(0)}\ecom
      \end{equation}
  where
      \begin{equation}  
        \Im{G_a^R(0)}=-\frac{\ell_B\Gamma_a(\edim=0,B)}{(E_a-\mdim)^2+(\ell_B\Gamma_a(\edim=0,B))^2}\edot
      \end{equation}
  In high magnetic fields $\mdim\to\ddim$, so we will use $\mdim\approx\ddim$. Using the high magnetic field limit of the scattering rate derived in Sec. \ref{sec:gamma} we see that in both cases the scattering rate is a power function $\Gamma\propto \ell_B^q$ ($q=2$ for $\ddim=0$ and $q=4$ for $\ddim\neq0$) and $\ell_B\Gamma\to0$. For $n>0$ Landau levels, $|E_a-\mdim|>0$ thus $C_{ab}\propto \ell_B^{(2q+2)}$ and as a consequence $\sigma_{xx}\propto \ell_B^{2q+1}$. The case of $n=0$ is more delicate since $|E_0-\mdim|\geq0$. In this case the limit gives a Dirac delta for the imaginary part of the Green's function $\Im{G_a^R(0)}\sim -\pi\delta(E_a-\mdim)$. As a function of $\pdim_z$ this becomes:
\begin{equation}
\label{eq:deltajac}
-\pi\delta(E_a-\mdim) = -\pi\frac{\sqrt{\mdim-\ddim}}{\mdim}\delta(\pdim_z\pm\sqrt{\mdim-\ddim})~\edot
\end{equation}
In the case of $\ddim=0$ this has no magnetic field dependence and thus $C_{ab}\propto\ell_B^{q+1}$ leading to $\sigma_{xx}\propto\ell_B^{q}$. But in the case of $\ddim\neq0$ Eq. (\ref{eq:deltajac}) is proportional to $B^{-1}$ and thus $\sigma_{xx}\propto\ell_B^{q+2}$. Since the $n=0$ case decays the least rapidly it will be the dominant at high magnetic fields so the overall magnetic field dependence of the conductivity becomes: $\sigma_{xx}\propto B^{-1}$ for $\Delta=0$ and $\sigma_{xx} \propto B^{-3}$ for $\Delta\neq0$.

\begin{figure}
      \includegraphics[width=.45\textwidth]{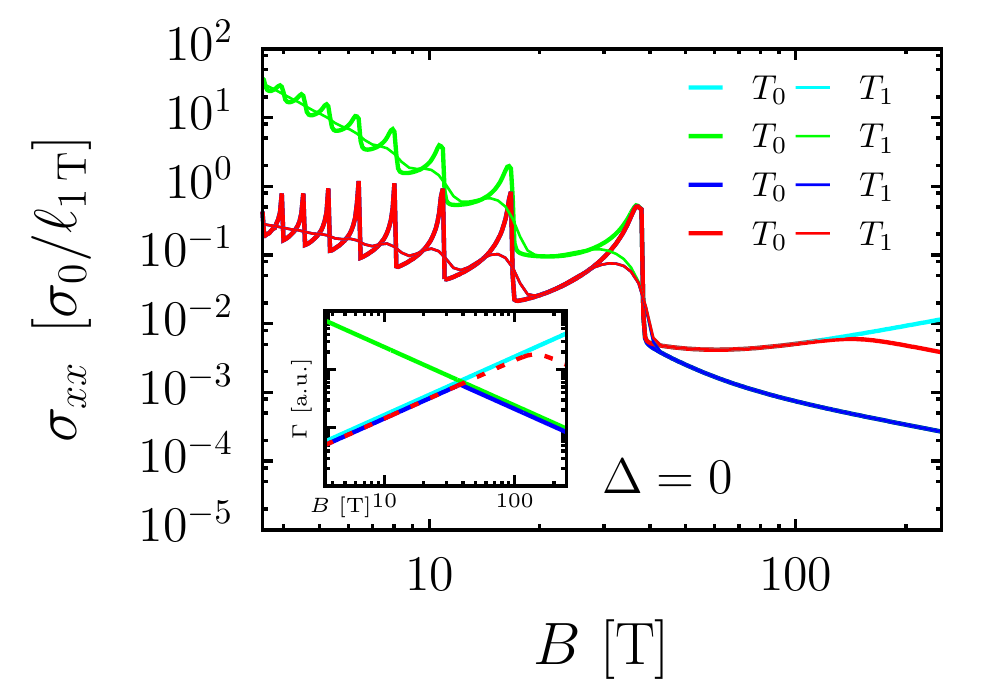}
      \vspace{-20pt}
      \includegraphics[width=.45\textwidth]{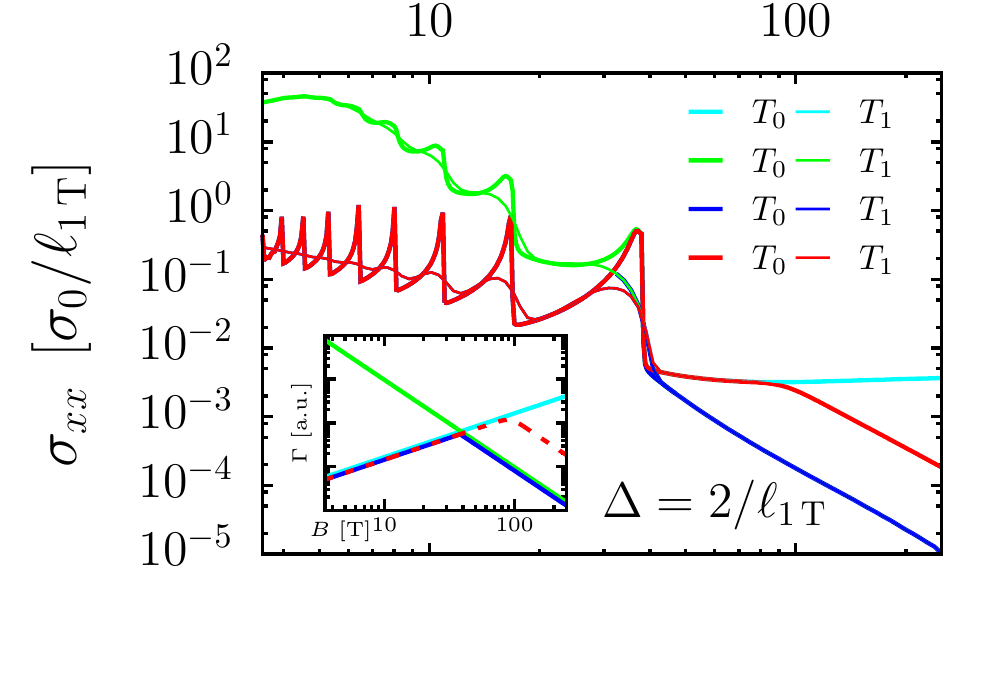}
      \caption{\label{fig:condxx} Transverse diagonal conductivity $\sigma_{xx}$ calculated from Eq. (\ref{eq:sigxx}) as a function of magnetic field at different temperatures. $\Delta=0$ (top plot) and $\Delta=2/\ell_{\SI{1}{\tesla}}$ (bottom plot). The scattering rate is chosen phenomenologically based on the numerical results in Fig. \ref{fig:Gamma_B}. The inset shows the scattering rates used (no Landau level dependence is assumed). $T_0=\SI{0}{\kelvin}$, $T_1=\SI{50}{\kelvin}$, the density of charge carriers is $n_e=10^{18}\si{\per\cubic\centi\metre}$ and $\sigma_0/\ell_{\SI{1}{\tesla}}\approx \SI{15}{\per\ohm\per\centi\metre}$.}
    \end{figure}

Next, we calculate $\sigma_{xx}$ assuming several choices of magnetic field dependence of the scattering rate in order to clarify its effects on $\sigma_{xx}$. The scattering rate is assumed to be independent of Landau levels and other variables except the magnetic field. The obtained results of $\sigma_{xx}$ for $\Delta=0$ and $\Delta=2/\ell_{\SI{1}{\tesla}}$ are shown in Fig. \ref{fig:condxx}. The impurity density is chosen in a way that the ratio of $\sigma_{xx}$ to $\sigma_{xy}$ in our results is similar to the experimental results\cite{Liang2015}.

When we assume that the scattering rate has the same magnetic-field dependence as described in Sec. \ref{sec:gamma} (i.e., $\Gamma\propto B^{-1}$ for $\Delta=0$ and $\Gamma \propto B^{-2}$ for $\Delta \neq 0$, green lines in the insets of Fig. \ref{fig:condxx}), the analytic behaviors in high magnetic fields are reproduced. (Note that the green lines in the main figures of Fig. \ref{fig:condxx} overlap with blue lines in the high field region.) However, in the low field region, we get a faster decrease than $B^{-1}$. This scattering rate is the same as used by Abrikosov\cite{Abrikosov1998}.
 
On the other hand, when we assume $\Gamma\propto B$ (cyan lines in Fig. \ref{fig:condxx}), we obtain $\sigma_{xx}\propto B^{-1}$ in the low field region. However, in this case, the analytic behaviors in high magnetic fields are not reproduced.

As shown in Fig. \ref{fig:Gamma_B} in Sec. \ref{sec:gamma}, the numerically obtained scattering rate roughly behaves as $\Gamma\propto B$ in the low field region, and $\Gamma\propto B^{-1}$ for $\Delta=0$ and $\Gamma \propto B^{-2}$ for $\Delta \ne 0$ in the high field region. Therefore, we connect these dependencies phenomenologically as shown with blue and red curves in Fig. \ref{fig:condxx}. In these cases, we obtain a $\sigma_{xx}\propto B^{-1}$ background with SdH oscillations superimposed in all the magnetic field region for the $\Delta=0$ case, while $\sigma_{xx}\propto B^{-1}$ in the low field region and $\sigma_{xx}\propto B^{-3}$ in the high field region for $\Delta\ne 0$ case. As shown in Fig. \ref{fig:Gamma_B}, there is no significant difference between $\Gamma_0$ for $\Delta=0$ and $\Gamma_0$ for $\Delta\ne 0$ in the low field region. The conductivity also behaves similarly, and the two cases behave differently only in the quantum limit where only the lowest Landau level is important.

About the temperature dependence, we can see it is negligible at high fields. This is because the temperature is only present in the $T\ell_B$ combination which goes to zero as the magnetic field gets higher. The effect of temperature is the suppression of the SdH oscillations. 

\begin{figure}
      \includegraphics[width=.45\textwidth]{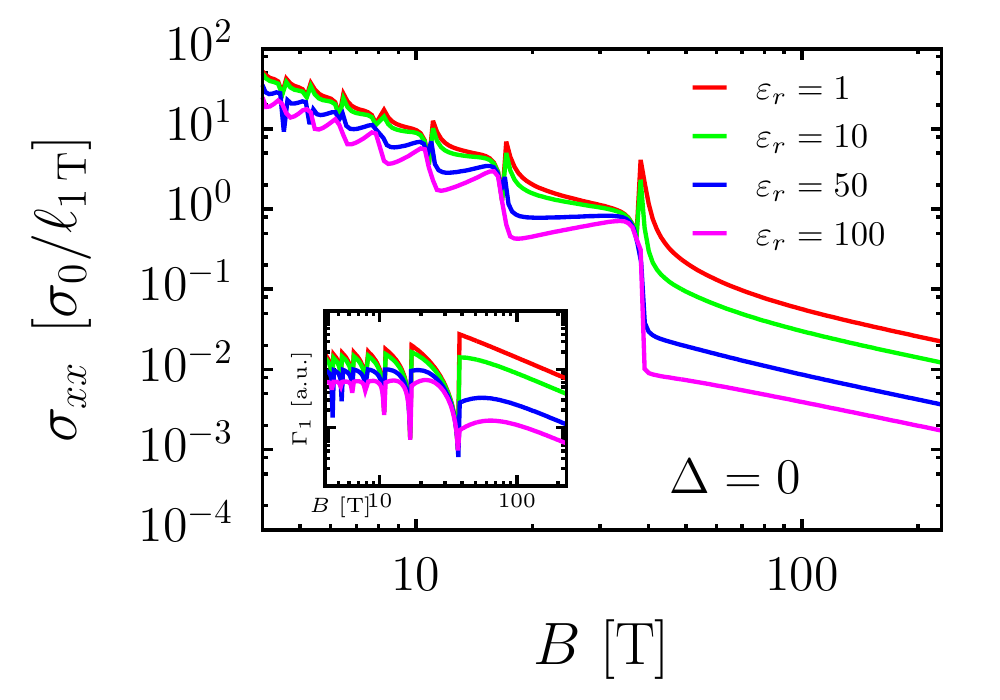}
      \vspace{-20pt}
      \includegraphics[width=.45\textwidth]{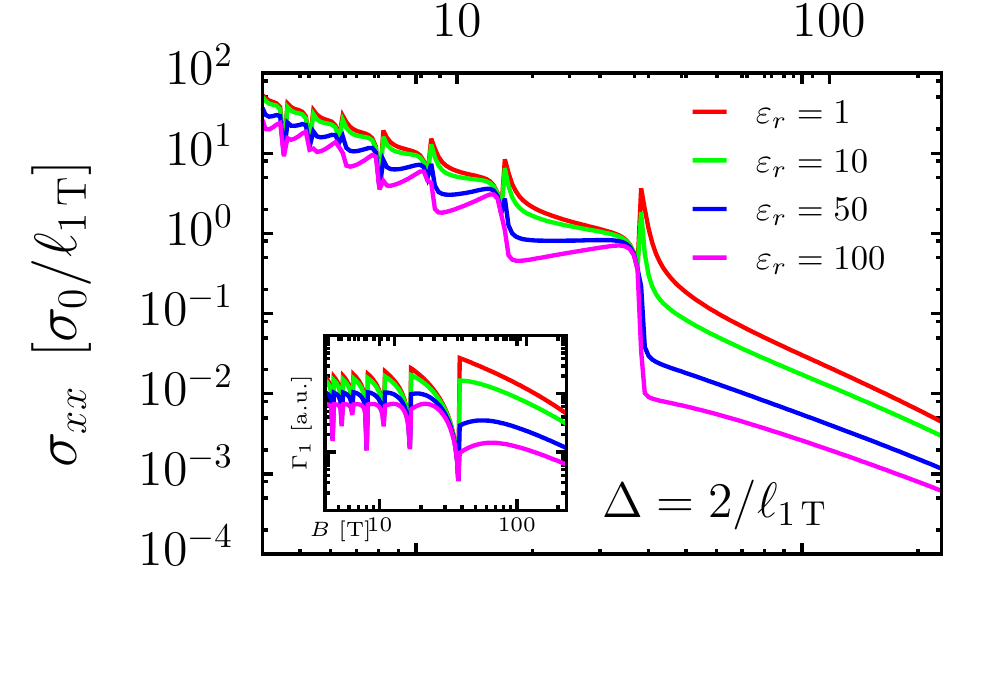}
      \caption{\label{fig:condxx_kap} Transverse diagonal conductivity $\sigma_{xx}$ calculated from Eq. (\ref{eq:sigxx}) as a function of magnetic field at zero temperature. $\Delta=0$ (top plot) and $\Delta=2/\ell_{\SI{1}{\tesla}}$ (bottom plot). The scattering rate is calculated using Eq. (\ref{eq:gam}) using screening wavenumbers calculated through Eq. (\ref{eq:wave}). The inset figure shows the scattering rates used (n=1 Landau level). The density of charge carriers is $n_e=10^{18}\si{\per\cubic\centi\metre}$ and $\sigma_0/\ell_{\SI{1}{\tesla}}\approx \SI{15}{\per\ohm\per\centi\metre}$.}
    \end{figure}

  A more precise numerical result can be achieved using the scattering rate calculated from Eq. (\ref{eq:gam}). However, the exact numerical integration of $\Gamma$ is a very heavy calculation. Therefore, we assume that the scattering rate is independent of momentum and energy ($\pdim_z=0$ and $\edim=0$) and only the Landau level dependence and magnetic field dependence are kept. For the strength of the interactions we assume $u_e=u_i=e^2/\varepsilon_0\varepsilon_r$ considering different relative permittivities. The results for both the massless and massive cases are shown in Fig. \ref{fig:condxx_kap}.

  In the high magnetic field region, we recover the magnetic field dependencies discussed above. In the low field region, $\sigma_{xx}\propto B^{-5/3}$ and the effect of the first Landau level appears as a very strong jump similarly to what was found in Ref. \onlinecite{Xiao2017}. We see that changing the relative permittivity changes the height of this jump. In the inset, we also show $\Gamma_1$, since this determines mainly the conductivity in high fields. In this system higher scattering rate means higher conductivity (contrary to normal system where the opposite is true). This means that if we increase the density of impurities the conductivity is also increased.

  \subsection{Magnetoresistance}
  The longitudinal resistivity is calculated as:
  \begin{equation}
    \label{eq:rhoxx}
    \varrho_{xx}=\frac{\sigma_{xx}}{\sigma_{xx}^2+\sigma_{xy}^2}\edot
  \end{equation} 

    \begin{figure}
      \includegraphics[width=.45\textwidth]{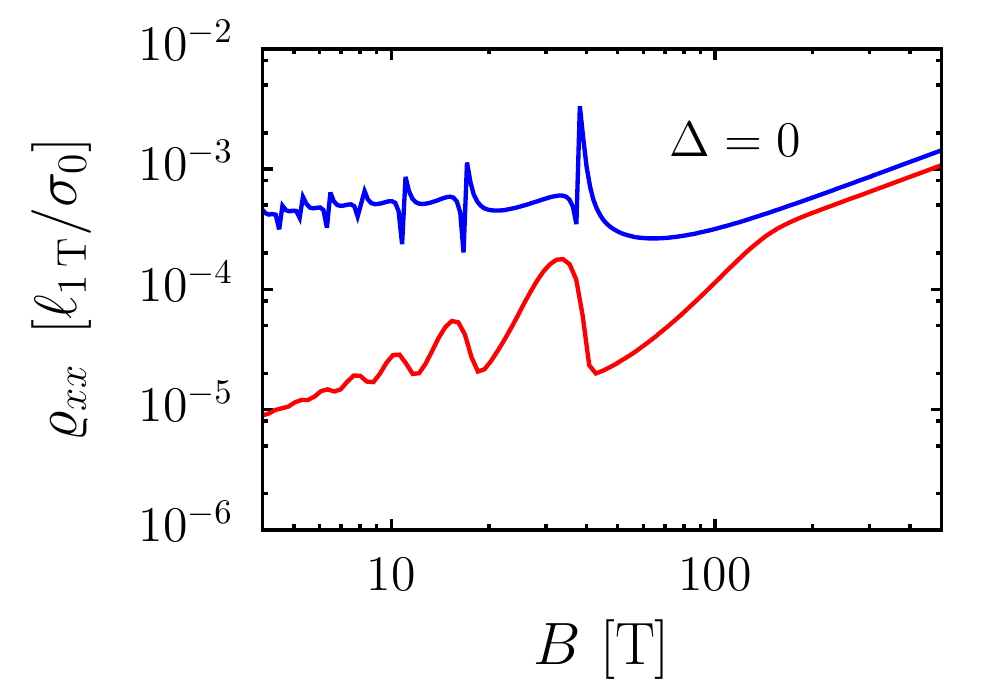}
      \vspace{-20pt}
      \includegraphics[width=.45\textwidth]{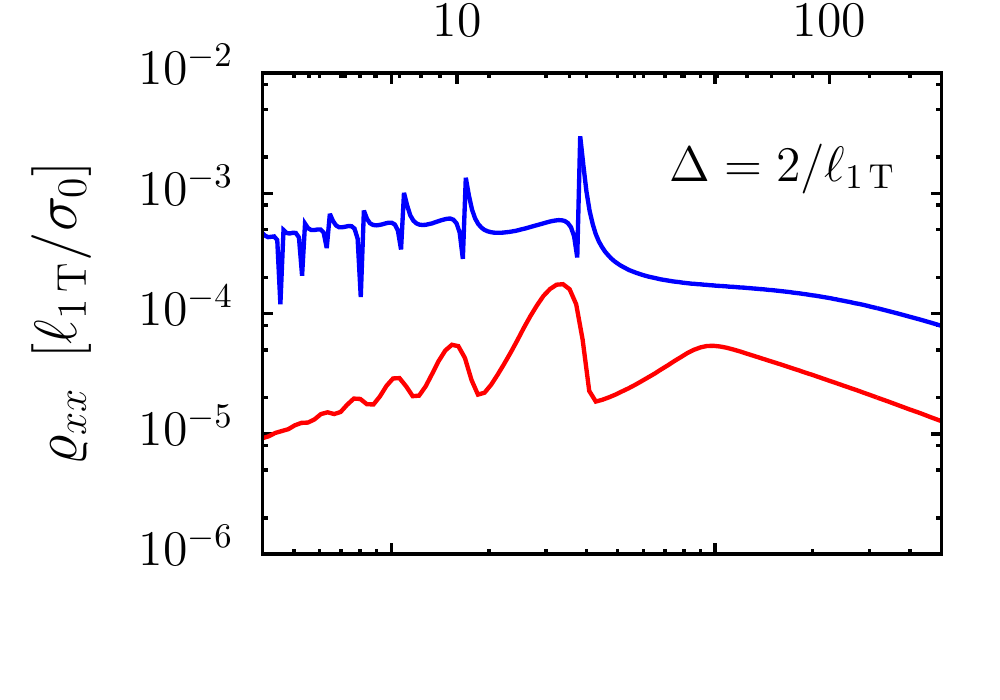}
      \caption{\label{fig:mag} Magnetoresistance $\varrho_{xx}$ calculated from Eq. (\ref{eq:rhoxx}) as a function of magnetic field. $\Delta=0$ (top panel) and $\Delta=2/\ell_{\SI{1}{\tesla}}$ (bottom panel). The resistivity calculated from the phenomenological result (red line) and the resistivity calculated microscopically using the first Born approximation with $\varepsilon=1$ (blue). The density of charge carriers is $n_e=10^{18}\si{\per\cubic\centi\metre}$ and $\sigma_0/\ell_{\SI{1}{\tesla}}\approx \SI{15}{\per\ohm\per\centi\metre}$.}
    \end{figure}

First, we discuss the magnetoresistance calculated from the phenomenological scattering rate represented by the red line in the insets of Fig. \ref{fig:condxx}. In this case, the obtained $\sigma_{xx}$ is proportional to $B^{-1}$ with SdH oscillations for the $\Delta=0$ case. The magnetoresistance calculated in Eq. (\ref{eq:rhoxx}) becomes $\rho_{xx} \propto B$ since both $\sigma_{xx}$ and $\sigma_{xy}$ are proportional to $B^{-1}$. This is shown with the red line in the top panel of Fig. \ref{fig:mag}. For the case of finite $\Delta$ (bottom panel), the lower field region (oscillating region) behaves similarly to the massless case as explained previously. The main difference is at high fields at the quantum limit. Since $\sigma_{xx} \propto B^{-3}$ the resistivity will be $\rho_{xx} \propto B^{-1}$ at high fields. This means that after the initial increase the magnetoresistance decreases at higher magnetic fields. However, we note that these results depend on the ratio of Hall conductivity and diagonal conductivity. We use that the diagonal component is smaller than the Hall conductivity. This is an experimentally reasonable assumption also done in Ref. 29. 

The magnetoresistances calculated using the numerically calculated scattering rates (corresponding to the case with $\varepsilon_r =1$ of Fig. \ref{fig:condxx_kap}) are shown with blue lines in Fig. \ref{fig:mag}. In the high field region, they behave similarly to those calculated using the phenomenological scattering rates. However, the low field behavior depends on the exact number of impurities, since the $B$ dependence of the conductivity is no longer $B^{-1}$ as shown in Fig. \ref{fig:condxx_kap}. 
\section{Summary and Discussions}

We studied the $4\times4$ massive Dirac Hamiltonian in a constant magnetic field which can be used as a simple continuum model for the gapped Dirac semimetals. This model shows certain similarities to the $2\times2$ Weyl Hamiltonian\cite{Abrikosov1998,Xiao2017} (i.e. massless case), but it contains several crucial differences.

The chemical potential was calculated implicitly fixing carrier density as a function of the magnetic field. For the gapless case we recover the result obtained in Ref. \onlinecite{Xiao2017}. We show that for $B\to\infty$ $\mu\to\Delta$. This behavior causes an important difference between the massive and massless case in both the scattering rate and the conductivity.

As we have seen in Sec. \ref{sec:gamma} the choice of screening wavenumber in the impurity potential greatly affects the magnetic field dependence of the scattering rate. We have calculated the screening caused by the electron-electron interaction through the random phase approximation (RPA). With this for high magnetic fields we have shown that the screening wavenumber increases as $\kappa^2\propto B$ for $\Delta=0$ and $\kappa^2\propto B^2$ for $\Delta\neq0$. Using the first Born approximation we have studied the scattering rate. For high magnetic fields $\Gamma\propto B^{-1}$ for $\Delta=0$ and $\Gamma\propto B^{-2}$ for $\Delta\neq0$. 

The Hall conductivity is shown to be inversely proportional to the magnetic field in the case of no impurities. In the case of a single Weyl node in Ref. \onlinecite{Xiao2017} Xiao et al. observed a small deviation from this behavior after the first Landau level crosses the chemical potential. In our case the symmetry in the quantum numbers and the explicit form of the eigenfunctions lead to the exact result in which $\sigma_{xy}$ behaves completely classically.

For calculating the diagonal transverse conductivity we discussed several possible magnetic field dependencies for the scattering rate. We have seen that the overall magnetoresistance is very sensitive to this choice. For high magnetic fields we show that $\sigma_{xx}\propto B^{-1}$ for $\Delta=0$ and $\sigma_{xx}\propto B^{-3}$ for $\Delta\neq0$. For lower fields at the oscillating region the massive and massless cases behave very similarly. It is shown that if the scattering rate is proportional to the magnetic field in this region, the conductivity becomes inversely proportional to $B$. We have seen that the temperature dependence is not so relevant. It decreases the SdH oscillations but it does not affect the overall magnetic field dependence. This is consistent with experimental results\cite{Narayanan2015,Niemann2017,He2014,Feng2015}. The temperature dependence in the experimental results is mainly caused by the normalization using the zero field conductivity (which is strongly temperature dependent).

For the magnetoresistance we recover the linear dependence for $\Delta=0$ at high fields. For the $\Delta\neq0$ case we see a decrease in the magnetoresistance as $\varrho_{xx}\propto B^{-1}$ at high fields. The main difference comes from the bottom of the lowest Landau level. In the massless case this level is linear and gapless, while in the massive case it is quadratic and gapped. At high magnetic fields when the chemical potential is close to the bottom of the lowest Landau level the difference becomes relevant. At low fields, $\varrho_{xx}$ for the massive and massless cases behave very similarly. In addition to the SdH oscillations, $\varrho_{xx}$ is proportional to $B$ if we assume a phenomenological scattering rate as $\Gamma\propto B$. On the other hand, if we use the scattering rate calculated from the Born approximation we get $\varrho_{xx}\propto B^{1/3}$ as in Ref. \onlinecite{Xiao2017}. Experimentally\cite{Canuto1970,Kaminker1981,Abrikosov1998,Klier2015a,Xiao2017,Klier2017a,Wang2018,Suetsugu2018}, the results are more consistent with the phenomenological case.

In Fig. \ref{fig:mag}, we showed an example of the behavior of $\varrho_{xx}$. This behavior depends on the choices of the carrier density $n_e$, the mass term $\Delta$, and the ratio between $\sigma_{xx}$ and $\sigma_{xy}$ which originates from the magnitude of $\Gamma$. As we can see in Fig. \ref{fig:mag} for the case of $\Delta\neq 0$, $\rho_{xx}$ changes its behavior from $\propto B$ in lower fields to $\propto B^{-1}$ in higher fields. The magnetic field at which this crossover occurs depends on the choice of $n_e$, $\Delta$, and $\Gamma$. The crossover magnetic field increases when $\Delta$ decreases. On the other hand, when $n_e$ becomes larger, the quantum limit occurs at a higher field and thus the crossover field also becomes higher. In experimental results for massive Dirac electrons only the linear behavior\cite{Suetsugu2018} is seen. This is consistent with our result, since the effective mass is small and the highest magnetic field used in the experiment is not high enough to get to the quantum limit.

It is natural to assume that the massive relativistic electron gas behaves similarly to the nonrelativistic electron gas at low energies. In the case of the nonrelativistic electron gas the diagonal conductivity is expected to saturate at high fields\cite{Abrikosov1988}. We do not see this behavior in the case of the gapped Dirac semimetal. The reason is because the two systems are similar only in the quantum limit when only the lowest Landau level is important. The nonrelativistic electron systems usually have a very high Fermi energy and the saturation is only valid when the chemical potential is high.

Finally, let us discuss the possible improvements of the present calculation. An important effect that was neglected in our formalism is the broadening of the density of states due to disorder. If we have used the impurity Green's function self-consistently the divergent peaks in the density of states at the bottom of the Landau levels would have been suppressed. As a consequence the oscillating behavior in both the chemical potential and the screening wavelength would have been modified. Also, because of self-consistency the scattering rate and conductivity would have been affected. Since the number of impurities is assumed to be small and since we are far from the charge neutrality point, similarly to Ref. \onlinecite{Xiao2017} we expect that this effect will not change the qualitative behaviors described in the paper.

 Using the static and long-wave limit in the RPA is the simplest way to include the electron-electron interaction for the screening. An improvement would be to take into account the frequency and momentum dependence of the polarization function and thus the screening wavenumber. The assumption of a simple screened Coulomb potential might not be sufficient to describe the effect properly.

The proper evaluation of the transport scattering rate through the vertex correction would further improve our calculation. In this paper we assumed that the approximative results obtained for the Weyl Hamiltonian\cite{Klier2015a} hold in our case as well. This should be revisited in greater detail.

 At lower fields we see very strong oscillations in the scattering rate, which are caused by the simple Born approximation. An important improvement would be the self consistent Born approximation. This is numerically very challenging.


\begin{acknowledgments}
We thank H.\ Matsuura, and H.\ Maebashi for fruitful discussions. This work was supported by a Grant-in-Aid for Scientific Research (B) on \lq\lq Multiband effects in magnetic responses and transport properties'' (No. 18H01162) from the MEXT of the Japanese Government. 
\end{acknowledgments}

\bibliography{bibliography.bib}

\end{document}